# Modeling Pressure Induced Structural Modification of Armchair Single-Wall Nanotubes

Ali Nasir Imtani and V. K. Jindal[1*]

Department of Physics, Panjab University, Chandigarh-160014, India

Abstract

Based on the helical and rotational symmetries and Tersoff-Brenner potential with couple of modified parameters, we investigate the variation of bond length/lengths in equilibrium structure due to tube length as well as due to applied hydrostatic pressure for a series of high symmetry armchair (n,n) single-wall nanotubes having different radii. Assuming that two different bond lengths dictate the tube geometry, these are monitored as a function of radius. It turns out that one of these bond lengths is greater than bond length of graphite whereas other one was less than it. These deviations from graphite value appear to be related to the curvature-induced rehybridization of the carbon orbitals. Lengths are found to have very important effect on the values of both bond lengths. The results under hydrostatic pressure indicate many linear behaviors having different slopes in the values of bond lengths with increasing pressure leading to a pressure induced-phase transition. This behaviour is strongly dependent on the tube radius. We also calculate the bulk moduls for this structure which reflects clearly this behavior of armchair nanotubes and thus predicts mechanical resilience of nanotubes.

## I- Introduction

Carbon nanotubes have excited a considerable interest in the condensed-matter and materials research communities, and much experimental and theoretical work has been devoted to them as prototype of one-dimensional ordered systems with promising technological applications [1]. The structure of carbon nanotubes is qualitatively well known through the simple construction of rolling a perfect graphite sheet, where only one parameter is to be determined: the lattice parameter or a bond length. The symmetry of the tube is less restrictive than in graphite and several parameters are needed to determine completely the structure. Among other things, these parameters define the differences between equivalent bonds, which condition the position of Fermi surface in the conducting armchair tubes. It is very difficult to obtain direct experimental information for the structure, and very little theoretical information has been given so far [2,3,4]. The strong similarity of the chemistry of carbon nanotubes to graphite allows theoretical analysis to be done based on empirical methodologies imported from studies on graphite. The curvature of the tubes, however, disturbs the chemistry in a way that can cause the deviation from the graphite based description, especially for small radii tubes. Roberston *et al* [5] used first principle LDF methods to calculate the total energies for a series of high symmetry tubules (n,n). They found that the minimum energy structure of (5,5) tubule by direct minimization of the total energy gives a radius of 3.47Å with both types of carbon-carbon bonds assumed equal with a length of 1.44Å. Daniel *et al* [1] have also reported the results based on two bond lengths using *ab initio* calculations. They found that both bond lengths have values grater than the value of bond length of graphite. Gao *et al* [22] have studied the energy, structure, mechanical and vibrational properties of armchair(n,n), zigzag(n.0) and chiral (2n,n) single-wall nonotubes of very short length (e.g. (10,10) tubes having 40 atoms) using an accutate interaction potential derived from quantum mechanics. For each form they studied two sets of initial structure, (1) perfect circular cross section and (2) elongated or collapsed cross section. They found that below 10 and 30Å both circular and collapse forms are possible, the circular cross section SWNT's are energetically favorable and beyond 30Å the collapsed form become favorable for all three types of SWNT.

---

[1] Author with whom correspondence be made, e-mail: jindal@pu.ac.in

Raman experiments [6] offer valuable information for the vibrations of ropes of single-wall armchair tubes. Resonant Raman scattering [7] has been used to measure the vibrations stemming from tubes of different diameters by looking at the $A_{1g}$ breathing mode that exhibit a strong dependence on the diameter. This therefore provides a technique to experimentally measure the effect of pressure on the tube.

Many studies [8-13] have been made about single-wall nanotubes under hydrostatic pressure. Venkateswaran *et al* [8] who examine the pressure dependence of the radial and tangential vibrational modes observe that the radial mode intensity vanished beyond 1.5GPa, suggesting that the disappearance of the radial mode intensity was due to polygonaization of nanotubes under pressure. Chesnokov *et al* [9] found an exceptionally large and reversible volume loss of SWNT's bundles under pressure in the 0-27 kbar range. The pressure-induced structural changes of single-wall carbon nanotubes organized in two dimensional crystalline bundles were studied [10]. They found a progressive deformation of the tube section from circular to hexagonal, in addition to van der Waals compression. Wu *et al*[11] have used first-principle quantum transport calculations aided by molecular dynamics simulation and continuum mechanics analysis to investigate the electronic transport properties of SWNT under hydrostatic pressure, and demonstrated that a reversible pressure induced shape transition for armchair SWNT's, which is turn induces a reversible electrical transition from metal to semiconductor. They found for (10,10) SWNT that the first transformation of tube from a circular to elliptical shape takes place at a critical transition pressure of 1.55GPa, and then from an ellipse to a dumbbell at pressure of 1.75GPa. Elliott *et al* [12] used classical molecular dynamics to examine the mechanical collapse and explore the response of diameter and chirality of nanotubes. They demonstrated that the single-wall carbon nanotubes bundles collapse under hydrostatic pressure and then obtained the collapse pressures as a function of a nanotube diameter and independent of the nanotube chirality. Jie Tang *et al* [13] used x-ray diffraction experiment under hydrostatic pressure to measure the volume compressibility of single-walled carbon nanotube of 14Å diameter. They found it equal to $0.024 GPa^{-1}$ and the single-wall nanotube polygonized when they form bundles of hexagonal closed-packed structure and the inter-tubular gap is smaller than the equilibrium spacing of graphite.

In order to have an insight in to such pressure induced structural transformation, it is preferable to have a detailed study based on established model potential. Further, it is necessary to understand the behavior of bond lengths under pressure for which no detailed study exist. This paper reports results on armchair carbon nanotubes structure, keeping there objective in mind. After giving theoretical preliminaries, we describe numerical procedure in the following. The results are discussed and concluded.

## II Theoretical Preliminaries

**A-The helical and rotational symmetries**

We can visualize an infinite tube as a conformal mapping of a two-dimensional honeycomb lattice to the surface of a cylinder that is subject to periodic boundaries both around the cylinder and along its axis. Atypical tube and a sheet are shown in Figure 1, where we also show the bond lengths $b_1$ and $b_2$. The bond length $b_1$ is perpendicular to the tube axis whereas $b_2$ make an angle with it. In this paper, we assume that the cross section of the SWNT remain circular. The helical and rotational symmetries [17] are used in this study to construct a high symmetry armchair (n,n) SWNT's (with $\theta = 30^o$). This is done by first mapping the two atoms in the [0,0] unit cell to the surface of cylindrical shape. The first atom is mapped to an arbitrary point on the cylindrical surface (e.q., $(R,0,0)$), where $R$ the tube radius in terms of bond lengths $b_1$ and $b_2$ and the position of the second atom is found by rotating this point by $\phi = 2\pi/3n$ (radian) about the cylinder axis. These first two atoms can be used to locate $2(n-1)$ additional atoms on the cylindrical surface by $(n-1)$ successive $2\pi/n$ rotations about the cylinder axis. Altogether, these 2n atoms complete the specification of the helical motif which corresponds to an area on the cylindrical surface. This helical can then be used to tile the reminder of the tubule by repeated operation of a single screw operation $S(h,\alpha)$ representing a translation $h$ unit along the cylinder axis and rotation $\alpha$ about this axis, where $h = \sqrt{3}b_2/2$ (unit) and $\alpha = 2\pi/n$ (radian), where $b_2$ is shown in Figure 1. If we apply the full helical motif, then the entire structure of armchair SWNT is generated. This structure provides atom position of all atoms in terms of bond length. The bond length is determined by minimization of the energy of the tube, assuming atoms interact via Tersoff-Brenner potential.

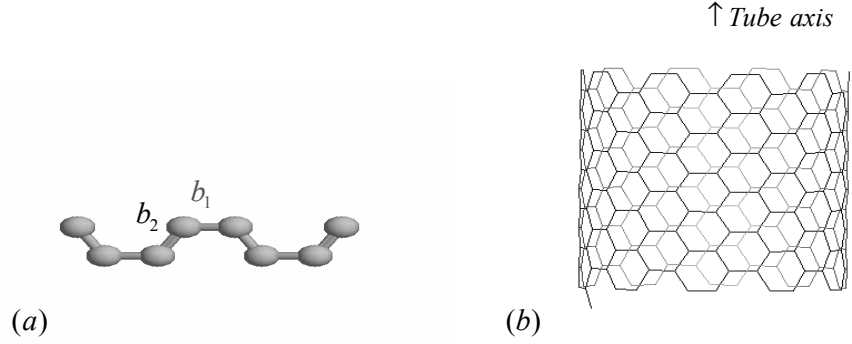

Figure 1: (a) A part of graphite sheet and (b) a schematic side view of armchair SWNT indicating two types of C-C bonds. These are labeled as $b_1$ and $b_2$.

**B-Tersoff-Brenner's potential**

One of the empirical interaction potentials of a covalent system is the Tersoff-Brenner Potential which has the following form [18,14]:

$$U = \sum_{ij} f_c(r_{ij})[Ae^{-\lambda_1 r_{ij}} - Bb_{ij}e^{-\lambda_2 r_{ij}}], \qquad 2.1$$

where $U$ is the total energy of the system. The indices $i$ and $j$ run over the atoms of the system, and $r_{ij}$ is the distance between atom $i$ and atom $j$. The first term represents a repulsive pair potential, which includes the orthogonalization energy when atomic wave functions overlap, and the second term represents an attractive pair potential associated with bonding. The term $f_c(r_{ij})$ is merely a smooth cutoff function, to limit the range of the potential in first neighbor shell. The function $b_{ij}$ gives non continuity to the potential and is a measure of the bond order. Further it depends upon local environment, and is assumed to be a monotonically decreasing function of the coordination of atoms $i$ and $j$. The terms which act to limit the range of interaction to the first neighbor shell are included in $b_{ij}$. The parameters $A, B, \lambda_1$ and $\lambda_2$ are all positive definite quantities. Here the cutoff function is simply taken as:

$$f_c(r) = \begin{cases} 1, & r < R-D \\ \frac{1}{2} - \frac{1}{2}\sin[\frac{\pi}{2}(r-R)/D], & R-D < r < R+D \\ 0, & r > R+D \end{cases} \qquad 2.2$$

which has continuous value and derivative for all r, and goes from 1 to 0 in a small range around $R$. $R$ and $D$ are chosen to include only the first-neighbor shell of most structure of interests. For graphite sheet the values of R and D are chosen as 1.95A° and 0.15A° respectively. $b_{ij}$ is taken to have the following form:

$$b_{ij} = (1 + \beta^{n1} \varsigma_{ij}^{n1})^{-\frac{1}{2n1}}$$

$$\varsigma_{ij} = \sum_{k \neq i,j} f_c(r_{ik}) g(\theta_{ijk}) \exp(\lambda_3^2 (r_{ij} - r_{ik})^2) \qquad 2.3$$

$$g(\theta_{ijk}) = 1 + \frac{c^2}{d^2} - \frac{c^2}{[d^2 + (h - \cos\theta_{ijk})^2]}$$

where $\theta_{ijk}$ is the bond angle between bonds $ij$ and $ik$. For graphite, $\theta_{ijk}$ is always $120^o$. In general, $\theta_{ijk}$ for SWNT are given $\theta_{ij} = \cos^{-1}\left(\frac{\vec{r}_i \cdot \vec{r}_j}{|\vec{r}_i||\vec{r}_j|}\right)$ with similar expression for $\theta_{ik}$. $\vec{r}_i, \vec{r}_j$ and $\vec{r}_k$ are the position vectors of atom $i$ and nearest neighbor atoms $j$ and $k$ on the cylinder surface. The bond-angle forces were introduced in the potential $\varsigma_{ij}$. The procedure used by Tersoff to calculate the parameters in the potential has been to fit the pair terms and an analytic expression for $b_{ij}$ to a number of properties of the diatomic and solid-state structure (e.g., bond energies and lengths, bulk moduli, vacancy formation energies, etc.). In graphite sheet each atom has local coordination of three (3, k=1, 3). This is because at this number of coordination the value of graphite sheet energy at minimum point and the nine parameters reported by Tersoff are [15]:

$A(eV) = 1.3036x10^3, B(eV) = 3.467x10^2, \lambda_1 = 3.4879 \text{Å}, \lambda_2 = 2.2119 \text{Å}, \beta = 1.5724x10^{-7}$, $n1 = 7.2751x10^{-1}, c = 3.8049x10^4, d = 4.384$, and $h = -5.7058x10^{-1}$.

They set the parameter $\lambda_3$ equal to zero.

The values of parameters $A$ and $B$ are found by Tersoff by adjusting the potential to bond length and energy of graphite values i.e. $1.46 \text{Å}$ [16], and $-7.4 eV/atom$ respectively. However, the experimental values of the bond length and energy for graphite are $1.42 \text{Å}$ and $-7.3756 eV/atom$ respectively. Since $A$ and $B$ chosen by Tersoff do not reproduce the bond length in graphite accurately, we so adjusted these two parameters to give closer agreement to bond length and energy to the experimental values of graphite values. The new values of these parameters we found are $A = 1206.7090 eV$ and $B = 315.96460 eV$. All other parameters we take them as above in our calculations will be performed for armchair (n,n) SWNT's, where n=3, 5, 7, 10, 15, 20, 30, 40 and 50 having different radii.

**III Numerical Procedure**

By using the helical and rotational symmetries and the Tersoff-Brenner potential with modified parameters $A$ and $B$ we investigated the variations of bond length associated with minimum energy of single-walled nanotubes. In general, two bond lengths $b_1$ and $b_2$ as shown in Figure 1 determines the structure of a carbon nanotube. However many times no distinction is made between these, and they are treated equal to each other, $b_1 = b_2 = b$. A given (n,n) SWNT can be constructed from the bond lengths. We allow $b$ to vary to obtain the position coordinates of a given length of an (n,n) tube and use Tersoff-Brenner potential to obtain its total energy. The bond length that results in minimum energy is numerically calculated. Similarly, we repeated this procedure to obtain a set of $b_1$ and $b_2$ that results in another minimum energy configuration as shown in Figure2 for armchair (7,7) tube. A final set of $b_1$ and $b_2$ was thus obtained by successive minimization procedure which eventually provide us with the two bond lengths. All calculations were made using this set of minimized bond lengths. The energy lowers by 0.14% by choosing $b_1 \neq b_2$ as compared to that when $b_1 = b_2 = b$ was chosen for (7,7) tube. Further a study of any variation with (n,n) SWNT's will provide information on dependence of bond length on curvature. For large radius carbon nanotubes, we would expect bond lengths to approach to that of graphite.

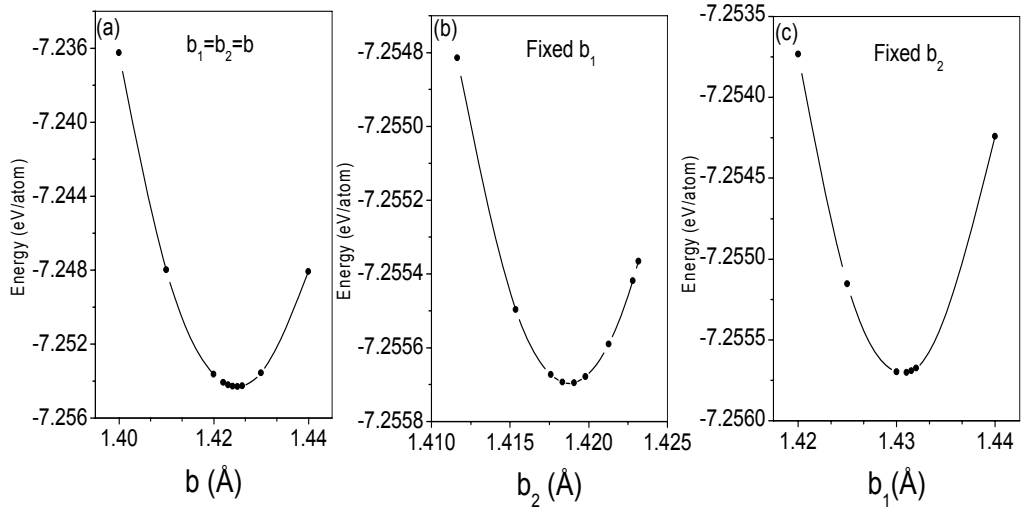

Figure 2: a, b and c show various intermediate steps leading to minimization of energy and corresponding bond lengths $b_1$ and $b_2$. The example here is for armchair (7,7) single-wall nanotube having 50 unit cell and 1400 atom.

After we demonstrated the successful procedure described above, then we applied it for armchair (n,n) SWNT's, where n=3, 5, 7, 10, 15, 20, 30, 40 and 50 having different radii with length determined by fixed number of unit cell, which makes the length of all tubes approximately equal, to investigate the equilibrium structure and length effects on the structure. These were then investigated further for changes under applied hydrostatic pressure.

**IV Results and Discussion**

**A Structure Properties**

We first study the equilibrium structure of armchair (n,n) SWNT's assuming that only the bond lengths are the variables; the bond angles are held fixed. By symmetry, for armchair tubes we found there are two inequivalant bonds. By following the minimization energy procedure as described in earlier section, we obtain the normalized values of these bonds lengths (i.e. $b_{1,2}/b_o$ where $b_o$ is the bond length in graphite sheet). There are plotted as a function of radius of tube in Figure 3a. In Table1 we present the results of our calculations for several armchair SWNT's. The differences between these bonds and graphite and between them are significant. Two effects are apparent: (i) one bond length $b_1$ is always greater than graphite value which is in agreement with those obtained by Daniel et al [1] using ab initio calculations, whereas other bond length $b_2$ is less than the graphite value in contrast with the calculations in the same reference. In our results of (5,5) tube we found it having the radius 3.43 Å and two bond lengths $b_1 = 1.44$ Å and $b_2 = 1.41774$ Å. The value of $b_1$ is in good agreement with the calculations done by Robertson et al [5] using first principle calculations to find the total minimized energy. They observe that the tubule have a radius 3.43Å and take bond lengths are equal with length 1.44Å. (ii) the behavior of both bond lengths is $b_1$ decreases and $b_2$ increases leading to approach graphite value when the radius of (n,n) tube increased. The difference between both bond lengths also decreases with increasing tube radius. Both effects can be easily understood in terms of rehybridization and weakening of the π bonds induced by the curvature [1]. The curvature energy can be calculated as $E_c = E_g - E_t$, where $E_g$ and $E_t$ are the energies per atom of graphite sheet and carbon nanotube under study. Table 1, Figure 3b and Figure 3c show the variations of energy and curvature energy as a function of the tubes radii. We observe that the radius is one important variable to decrease the curvature effect. It is however interesting to note (Figure 3c) that even (50,50) tube with radius 33.9238Å does not have vanished curvature effect.

Table 1: Armchair SWNT, radius, normalized bond Lengths, energy E(eV/atom) and curvature energy $E_c$ (eV/atom) of armchair SWNT's with length consisting (≈122.0Å) of 50 unit cells.

| SWNT | Radius(A°) | $b_1/b_o$ | $b_2/b_o$ | E | $E_C$ |
|---|---|---|---|---|---|
| (3,3) | 2.11134 | 1.0380 | 0.9983 | -6.961457 | 0.414142 |
| (5,5) | 3.43774 | 1.0140 | 0.9984 | -7.195806 | 0.179793 |
| (7,7) | 4.78276 | 1.0074 | 0.9993 | -7.255701 | 0.119898 |
| (10,10) | 6.80626 | 1.0038 | 0.9995 | -7.286506 | 0.089093 |
| (15,15) | 10.1914 | 1.0021 | 0.9998 | -7.302638 | 0.072961 |
| (20,20) | 13.5790 | 1.0014 | 0.9999 | -7.308231 | 0.067368 |
| (30,30) | 20.3543 | 1.0007 | 0.9999 | -7.312207 | 0.063392 |
| (40,40) | 27.1391 | 1.0007 | 0.9999 | -7.313597 | 0.062002 |
| (50,50) | 33.9238 | 1.0007 | 0.9999 | -7.314239 | 0.061360 |

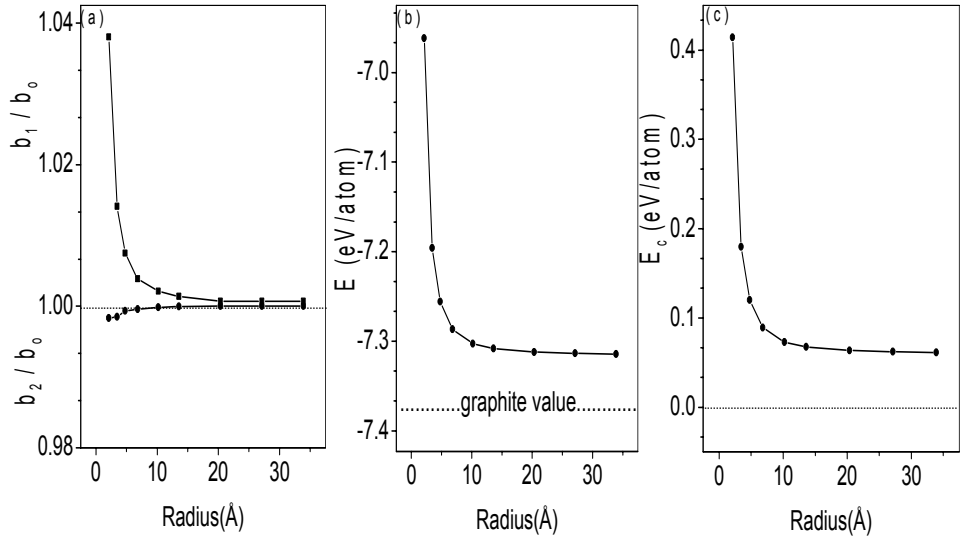

Figure 3: (a) Variations of the normalized bond lengths $b_1/b_0$, (b) energy $E$ and (c) curvature energy $E_c$ as a function of radius of tube for armchair (n,n) single-wall nanotubes, where n=3, 5, 7, 10, 15, 20, 30, 40 and 50. It is clear that as radius increases all microscopic parameters of nanotubes approximately approach to graphite values. Line draw is only a guide to the eye.

**B- Length effects**

As in section A, we investigated the variations of the both bond lengths with the radius of tube, we also investigate here in this section the effect of tube length on the values of both bond lengths. We choose for this aim three armchair (5,5), (15,15) and (30,30) single-wall nanotubes having small, medium and large radii respectively. We compare the structure of these tubes as a function of length aspect ratio, where length has been measured in unit of radius. Figure 4a shows the variations of both bond lengths (in unit of graphite value) for nanotubes plotted as a function of length to radius ratio. We observe that: the bond length $b_1$ decreases towards the value of bond length of graphite when the length of tubes increases, whereas the bond length $b_2$ increases toward the graphite value with increasing length of tube. The difference between both bond lengths with the tube length is very significant. For (5,5) tube having small radius, although the values of both bond lengths move toward graphite value but the difference between them is always significantly large, even for larger $L/R$. The major reason of this shift in the values of bond lengths of armchair carbon nanotubes from graphite value is because of the curvature. For reasonably long lengths of the tubes, $b_1/b_2$ approach a constant value which depends on the radius of the tube. We present this data on $b_1/b_2$ for these tubes in Table 2. We see that when the ($L/R \approx \infty$) the value of both bond lengths are close to each other as compared to when $L/R \approx 1$. Our calculations indicated as shown in Figure 4c that the curvature is responsible for the difference in $b_1$ and $b_2$ which may not approach graphite value on increasing length only. We do not expect that increase in length only will make $b_1$ close to $b_2$. As a result of our prenliminary calculations on zigzag and chairal tubes; the length of unit cell of armchair shorter than the uint cells in zigzag and chiral SWNT's, the length have slightly effect on the value of bond lengths. Similarly the length have important effect on the value of energy of nanotubes as we see from the Figure 4b that the increasing of tube length tends to change the value of its energy toward the graphite value by decreasing of the effect of curvature due to length of tube as shown in the same Figure.

Table 2: Radius and normalized bond lengths and bond lengths ratio in different values of length to radius ratio of (5,5),(10,10) and (30,30) tubes

| SWNT | Radius(Å) | $b_1/b_o$ | Radius(Å) | $b_1/b_o$ | $b_1/b_2$ | |
|---|---|---|---|---|---|---|
|  | $L/R \approx 1$ | | $L/R \approx \infty$ | | $L/R \approx 1$ | $L/R \approx \infty$ |
| (5,5) | 3.349234 | 1.0204 | 3.437746 | 1.01408 | 1.4490 | 1.01569 |
| (15,15) | 10.21297 | 1.0035 | 10.19041 | 1.00200 | 1.0056 | 1.00204 |
| (30,30) | 20.49756 | 1.0014 | 20.36864 | 1.00070 | 1.0025 | 1.00056 |

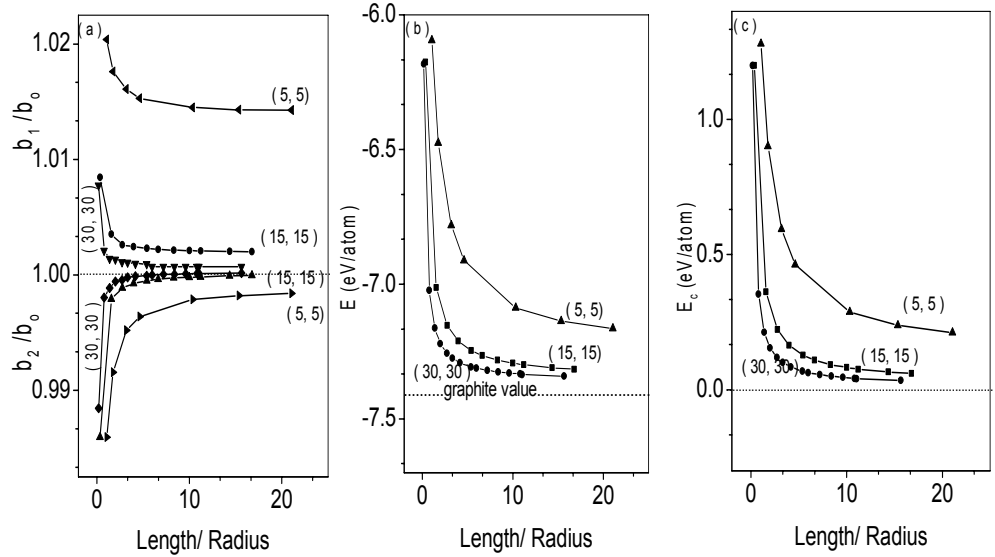

Figure 4: Microscopic parameters plotted as a function of the length to radius ratio for armchair (n,n) single-wall nanotubes (n=5, 15 and 30). (a) Normalized bond lengths $b_1/b_o$ and $b_2/b_o$. We see that as length of tube increases the bond lengths approach each other and to graphite value. (b) Energy $E$ and (c) curvature energy $E_c$. However (5,5) tube take stiffer $b_1/b_o$. Lines are only used as guide to the eye.

## C- Pressure effect

SWNT's are flexible fibers and can be 100 times stronger than steel [19]. Pressing on the tip of the nanotube will cause it to bend without damage to the tip or the whole SWNT. When the force is removed, the tip of nanotube recovers to its original shape [20]. Here in this section we study the pressure effects on the structure of armchair SWNT's.

An applied of a hydrostatic pressure P in axial and radial directions on armchair single-wall carbon nanotubes alters the total potential energy of tube and then from the first law of thermodynamics is written as:

$$U = U_o + P\Delta V, \qquad 2.4$$

where $\Delta V = V - V_o$ is the volume reduction, V the produced volume under applied the pressure P, $V_o$ and $U_o$ are the volume and energy at zero pressure.

The procedure used here is to investigate the structure of SWNT under pressure is that first we assume that only one bond length is reduced by some amount and let this new bond length $b_{1p}$ and then we search about the suitable value of the second bond length $b_{2p}$ by the same procedure which it used in this work to calculate the bond lengths and minimum energy at zero pressure. By successive minimization procedures, we calculate a set of $b_{ip}$ and $b_{2p}$ value that result in minimizes energy under some hydrostatic pressure P.

After we get the value of minimum energy U and the values of the new bond lengths under the pressure, the radius $R_p$ and length $L_p$ of nanotube become known. Thus, the volumes of the circular cylinder shape are $V = \pi R_p^2 L_p$ and $V_o = \pi R_o^2 L_o$, where $R_o$ and $L_o$ are the radius and length of nanotube at zero pressure. Pressure can then be calculated from these data, $P = -\dfrac{\Delta U}{\Delta V}$.

We have studied the pressure dependence structure for four armchairs SWNT's having different radii. Figure 5 shows the results of our calculations for (5,5), (10,10), (15,15) and (30,30) SWNT's under a hydrostatic pressure. Due to difference in bond lengths, hydrostatic pressure results in unequal compression in radial and axial direction.

The relation between variations of both bond lengths $b_1$ and $b_2$ and applied pressure appear to have linear behavior, though with different slopes as shown in Figure 5a for (10,10) tube. As can be observed from this Figure $b_1$ and $b_2$ has several steps of variation under application of pressure. All other tubes in this study have similar behavior. The extensive parameters of tube like length, radius, volume and energy are dependent on the two structural bond lengths. We plot these as a function of pressure in Figures 5b and 5c. Figure 5d show our results for the fractional change in the length and radius under the pressure for (10,10) tube. These results indicate that the carbon nanotube is extremely rigid along the tube axis than in the radial direction as proposed by Reich [21]. Bulk moduls can also be found from $B = -V\dfrac{dP}{dV}$, where V the sample volume and P applied pressure. The results of bulk moduls for these four tubes are listed in Table 3. One notices several critical pressure values that result in jumps in bulk moduli. The reason of present many ranges in values of pressure as we see in this Table, the behavior of bond lengths under pressure as disused above reflects in the behavior of bulk moduls also. For example the first range for (5,5) tube occur at pressure 1.85GPa, as compared to (10,10) tube the first range of pressure occur at 0.74GPa. As the radius increases as to (15,15) and (30,30) tubes the values of pressures are 0.398GPa and 0.08GPa, respectively. These values are in the same reduction percent of volume ratio ~ 99.3% for all tubes. Bulk moduls values are increasing with pressure that means the tubes become more solid under the pressure. In Figure 6, we show the dependence of bulk moduls with pressure for (10,10) tube as well as with the radius for these four tubes in this study. The tubes become easier to compress on increasing the radius. For very large radius, one can extrapolate $B$ to value $\approx 46.5 GPa$, which can be compared to the value of graphite. The major reason of the high values of pressure is that the changing of coordination number from three to seven at bond length $b_1$ equal to 1.165Å and to twelve at bond length $b_1$ equal to 1.12Å, where the repulsive force between atoms in the first shell increasing significantly and leads to resist any external force induced charge.

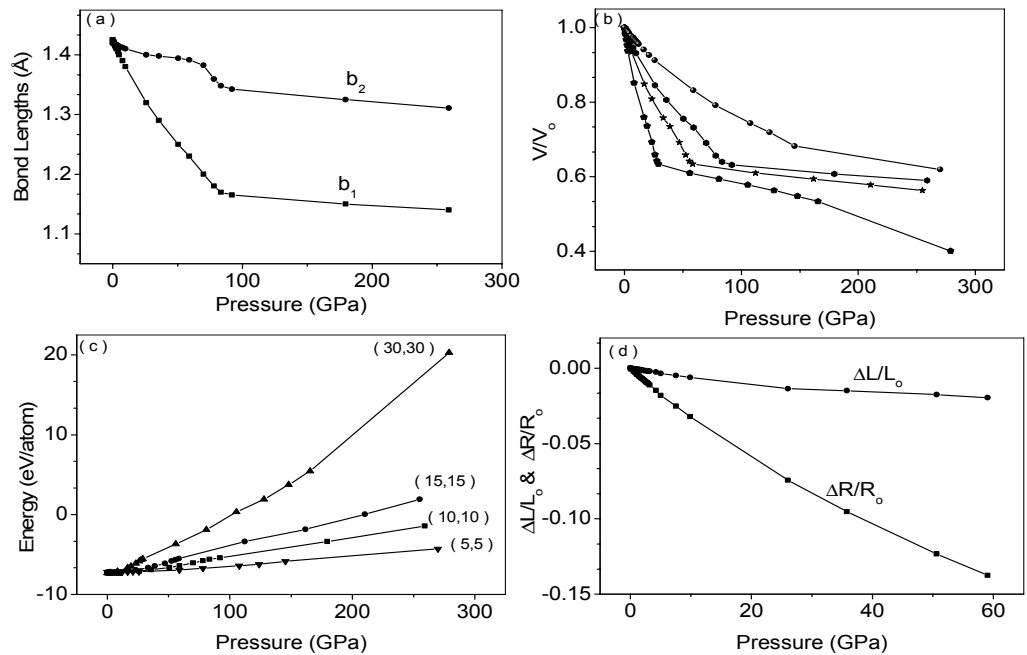

Figure 5: (a) Variations of bond lengths vs hydrostatic pressure for armchair (10,10) SWNT. (b) Volume ratio as a function of pressure of armchair (5,5), (10,10), (15,15) and (30,30) single-wall nanotubes. The value of pressure at the same volume ratio dependent on the radius of tube. (c) Energy with pressure shows that the tube with small radius is more resistance than tube with large radius. (d) The compression of tube along tube axis and in circumference directions under applied pressure for (10,10) SWNT to explain the solidity of tube along its axis and there are similar curves for another tubes.

Table 3: Hydrostatic pressure and bulk moduls for armchair (5,5), (10,10),(15,15) and (30,30) SWNT's.

| (5,5) | | | | | | |
|---|---|---|---|---|---|---|
| P(GPa) | 0.0 - 3.1 | 3.9-12.1 | 16.0 –77.9 | 107.0 –145.4 | | |
| B(GPa) | 262.30 | 273.34 | 411.25 | 616.19 | | |
| (10,10) | | | | | | |
| P(GPa) | 0.0–1.14 | 1.35-2.36 | 2.58 – 4.27 | 5.0 – 9.8 | 26.0-50.6 | 59.0-91.9 |
| B(GPa) | 133.45 | 134.46 | 140.80 | 164.20 | 272.6 | 297.0 |
| (15,15) | | | | | | |
| P(GPa) | 0.0-0.39 | 1.0-6.2 | 16.8-58.4 | | | |
| B(GPa) | 86.577 | 97.094 | 193.155 | | | |
| (30,30) | | | | | | |
| P(GPa) | 0.0-0.7 | 1.5-8.4 | 16.3-29.0 | | | |
| B(GPa) | 46.50 | 60.40 | 95.126 | | | |

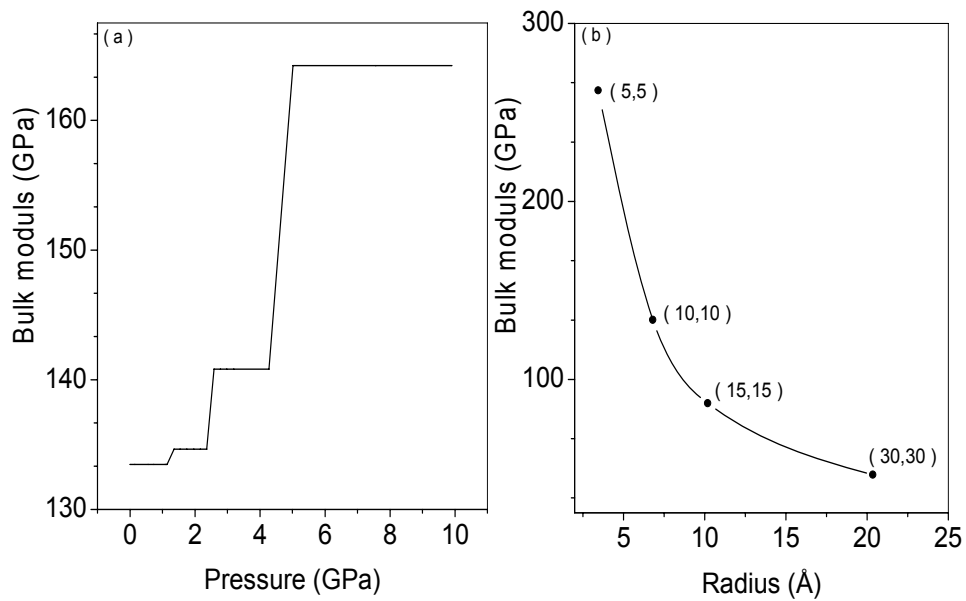

Figure 6: Variations of bulk moduls as a function of (a) applied pressure for (10,10) SWNT. We see the appearing of many ranges causing by the blinear behavior with different slopes of bond lengths under pressure and (b) tubes radii in the first range for (5,5), (10,10), (15,15) and (30,30). The smallest radius tube has the largest value of bulk moduls; more solidity than the larger radii tubes in this study.

# V Summary and Conclusion

This paper reports several interesting results on measurable properties of armchair SWNT's based on computation of two bond lengths. The methodology is simple and transparent. Using a well defined and well established intra-molcular potential, computational and analytical results based on minimum energy calculations give the two bond lengths. These are essential ingredients required for the evaluation of measurable properties. The whole calculation and procedure gives a lot of insight in to the mechanism of radius, length and pressure dependent results.

We have used the helical and rotational symmetries with a reliable Terssof-Brenner potential to investigate the variations of the values of bond lengths as a function of tube radius as well as a function of length of tube for series of armchair single-wall nanotubes with different radii using the procedure described in section III. After we get the structure (values of bond lengths) of armchair single-wall nanotubes at ambient conditions, we recalculate these under hydrostatic pressure. We chose four tubes having different radii to study the behavior of these bond lengths under pressure by using our procedure and the methodology described in section IV. The importance of this study is that all unique properties of nanotubes are depending on radius or chairlty of nanotubes. The radius of single-wall nanotubes is in terms of the bond lengths and the values of these bond lengths come out to be different from the value of the bond length of graphite sheet.

The central conclusion of this work is that the structure of armchair nanotubes have unequal bond lengths, one of these bonds having value more than graphite value whereas the other value being less that of graphite. These bond lengths are very sensitivite to the radius due to curvature effects as well as to the length of tube, especially, when the tube is short. The behavior of armchair single-wall nanotubes under hydrostatic pressure indicates these to be flexible tubes and the compression along the tube axis is less as compared to that along the circumference. Bulk modulii values are found to depend on the tube radius. Tubes with small radii are rigid having grater value of bulk modulii than the tubes with larger radii at the same applied pressure.

Therefore, effective theoretical calculations for electronic, mechanical and thermodynamical properties should be based on two bond lengths of carbon nanotubes in question, especially if the tubes are shorter in radius and length. We hope that this work provide enough motivation for more experimental work based on structural modification of armchair SWNT's of various lengths and under various hydrostatic pressures. Raman scattering around breathing mode can be useful for radius determination, whereas length may also be possible by small angle scattering techniques.